\documentclass[aps,pra,showpacs,eqsecnum,twocolumn]{revtex4}
\usepackage{graphicx}
\usepackage{amsmath,amssymb,amsthm,dsfont,bm}
\usepackage{color}
\usepackage{soul}

\begin{document}
\preprint{}
\title{Complementarity and the pathological statistics of the quantum impossible}
\author{Irene Bartolom\'e} 
\author{Alfredo Luis}
\email{alluis@fis.ucm.es}
\homepage{http://www.ucm.es/info/gioq}
\affiliation{Departamento de \'{O}ptica, Facultad de Ciencias
F\'{\i}sicas, Universidad Complutense, 28040 Madrid, Spain}
\date{\today}

\begin{abstract}
Standard quantum physics prevents  the existence of a joint statistics for complementary observables. Nevertheless, a  
joint distribution for complementary observables can be derived from their imperfect simultaneous measurement followed 
by a suitable data inversion. Quantumness is reflected then in the pathologies of the inferred distribution. We apply this 
program to the paradigmatic example of complementarity: the wave-particle duality in a Young interferometer.
\end{abstract}


\maketitle

\section{Introduction}
Complementarity is a distinguished quantum feature that precludes from the start the simultaneous exact observation of 
conjugate observables. This makes impossible even the mere conception of a joint probability distribution. But nothing prevents their 
less than perfect observation, providing us with an operational joint  distribution for complementary variables \cite{WMM}. Then, 
as far as we know all the details of our measuring scheme, we know the way this instrumental uncertainty has been added to each variable. 
Thus we can remove the effect of the measurement from both observables and get their exact distributions. This inversion can be 
then applied to the operational joint  distribution. In classical physics this program works giving their {\em bona fide} exact 
joint distribution \cite{AL15}. But in the quantum domain this is an attempt to obtain the quantum impossible, so quantum mechanics  
manifests in the form of pathological distributions, that nonetheless have perfect marginals for both observables. We may say that this 
is actually the hallmark of quantumness and the way nonclassical states are defined in quantum optics \cite{MWSZ}. 

We apply the above program to a seminal example of complementarity: the single-particle Young interferometer. The two conjugate 
observables are the slit crossed and the interference, or phase difference. Their joint observation will be allowed by marking the 
slit crossed in the polarization or spin state of the interfering particle. Then the interference is observed keeping track of the 
polarization/spin state when recording the interference. As mentioned above, this program requires that the path observation must 
be less than perfect so that the interference is not fully destroyed \cite{SIM}. 
 
 \section{Exact, unobserved statistics}
 
 We address here the basic definitions of  states and observables and their exact statistics before observation. Let us express the 
 state at the plane of the apertures of a Young interferometer by a two-dimensional complex vector $|\psi \rangle= ( \alpha, \beta )$, 
 where $(1,0)$ means particle in the upper aperture and $(0,1)$ particle at the lower aperture with $ | \alpha |^2+ |\beta |^2 = 1$.
 This is to say that we can represent the path observable by  the third Pauli  matrix $Z$, and denote as $z=1$ the particle in the 
 upper aperture and $z=-1$ as the particle in the lower aperture. The corresponding statistics are
 \begin{equation}
 P_Z (z=1) = | \alpha |^2 , \qquad  P_Z (z=-1) = | \beta |^2 ,
 \end{equation}
 or equivalently 
  \begin{equation}
  \label{PZe}
 P_Z (z=\pm 1) =  \frac{1}{2} \left (1 + z \langle Z \rangle \right ), \quad  \langle Z \rangle =  |\alpha |^2- | \beta |^2 .
 \end{equation}
  
 The conjugate observable is slightly more difficult to be described, as far as one expects interferograms with a continuous distribution 
 in the form of fringes. More rigorously the phase/interference observable should involve both the other two Pauli matrices $X$, and $Y$. 
 But for the sake of simplicity let us first properly represent phase/interference by just one of them, say $X$. Deep down, this corresponds 
 to say that $\langle Y \rangle =0$, or in any case, to represent phase difference by the unitary operator exponential of phase introduced 
 in Ref. \cite{PD}.

The statistics of $X$ is given by projection of the following vectors, expressed  in the same basis used up  to now: 
 \begin{equation}
 \label{ps}
 | x = \pm 1 \rangle = \frac{1}{\sqrt{2}} \left ( 1, \pm 1 \right ) ,
 \end{equation}
 and then 
 \begin{equation}
 \label{PXe}
 P_X (x) =  \frac{1}{2} \left (1 + x \langle X \rangle \right ), \quad  \langle X \rangle = \alpha \beta^\ast + \alpha^\ast \beta .
 \end{equation}
 
 In a later section we will consider the alternative approach where phase is represented by the positive operator valued measure given 
 by projection on the nonorthogonal phase states \cite{POVM}
 \begin{equation}
 | \phi \rangle = \frac{1}{\sqrt{2\pi}}  \left ( 1 , e^{i \phi} \right ) ,
 \end{equation}
 so that the exact phase distribution is 
 \begin{equation}
 P_\Phi (\phi) =  \frac{1}{2 \pi} \left (1 +  \cos \phi \langle X \rangle +   \sin \phi \langle Y \rangle \right ) ,
 \end{equation}
 where $ \langle Y \rangle = i (\alpha \beta^\ast - \alpha^\ast \beta)$.

 \section{Joint observation: discrete phase}
 
To perform a simultaneous observation of $X$ and $Z$ we must involve additional degrees of freedom. For example let us transfer path 
information to the polarization or spin state of the interfering particle. This can be easily achieved in practice with a half-wave plate in 
the case of photons, and a suitable arranged magnetic field in the case of massive particles. Le us describe the spin state by two 
orthogonal base vectors $| \rightarrow \rangle$ and $| \uparrow \rangle$. The particle is initially prepared in the state $| \rightarrow  \rangle$ 
and it continues in this same state after crossing the lower aperture, but changes its state to 
\begin{equation}
\cos \theta | \rightarrow \rangle + \sin \theta | \uparrow \rangle ,
\end{equation}
after crossing the upper aperture, where $\theta$ is some arbitrary angle. Thus, the complete state after this transfer 
is the path-spin entangled state 
\begin{equation}
| \tilde{\psi} \rangle = \left ( \alpha , 0 \right ) \left ( \cos \theta | \rightarrow \rangle + \sin \theta | \uparrow \rangle \right ) + 
 \left ( 0, \beta \right )  | \rightarrow \rangle .
\end{equation}   

The spin state is measured by projection on the following orthogonal states 
\begin{eqnarray}
& | \tilde{z} =1\rangle = \cos \vartheta | \rightarrow \rangle + \sin \vartheta | \uparrow \rangle , & \nonumber \\
 & & \\
 & | \tilde{z} = - 1 \rangle = - \sin \vartheta | \rightarrow \rangle + \cos \vartheta | \uparrow \rangle , & \nonumber 
\end{eqnarray}
where $\vartheta$ is an arbitrary angle. This can be easily achieved by a linear polarizer for photons and via an Stern-Gerlach 
apparatus for massive particles. On the other hand the interference is measured as before via protection on the states (\ref{ps}). 
 
Thus the statistics for the so constructed joint observation of interference and polarization is
 \begin{equation}
 \tilde{P} (x,z)  =  \left | \langle \tilde{z} | \langle x |   \tilde{\psi} \rangle \right |^2,
 \end{equation}
leading to 
\begin{equation}
 \tilde{P} (x,z)  = \frac{1}{2} \left [ \gamma_0 (z) + x \gamma_X (z) \langle X \rangle + z \gamma_Z (z) \langle Z \rangle \right ] ,
 \end{equation}
 where
\begin{eqnarray} 
\label{g}
& \gamma_0 (1) = \frac{1}{2} \left [ \cos^2 (\vartheta - \theta ) + \cos^2 \vartheta  \right ]  ,& \nonumber \\
& \gamma_0 (-1) = \frac{1}{2} \left [\sin^2 ( \vartheta - \theta ) + \sin^2 \vartheta \right ] , & \nonumber \\
& \gamma_X (1) = \cos (\vartheta - \theta) \cos \vartheta  ,& \nonumber  \\
& \gamma_X (-1) = \sin (\vartheta - \theta) \sin \vartheta , & \nonumber  \\
& \gamma_Z (1) =  \frac{1}{2} \left [ \cos^2 (\vartheta - \theta ) - \cos^2 \vartheta  \right ]  ,& \nonumber \\
& \gamma_Z (-1) =  \frac{1}{2} \left [ - \sin^2 (\vartheta - \theta ) + \sin^2 \vartheta  \right ]  ,&
  \end{eqnarray}
The corresponding marginals are:
\begin{equation}
\label{PXa}
\tilde{P}_X (x) = \frac{1}{2} \left ( 1 + x \cos \theta \langle X \rangle \right ) ,
 \end{equation}
and 
\begin{equation}
\label{PZa}
 \tilde{P}_Z (z)  = \gamma_0 (z) + z \gamma_Z (z) \langle Z \rangle .
 \end{equation}
  
 \section{Data inversion and impossible statistics: discrete phase}
It is possible to obtain the exact $P_A$ statistics (\ref{PZe}) and (\ref{PXe}) from the operational ones  $\tilde{P}_A$  in Eqs.  (\ref{PXa}) and (\ref{PZa}),
$A=X, Z$ in an extremely simple linear way as 
\begin{equation}
P_A (a) = \sum_{a^\prime} \mu_A (a, a^\prime ) \tilde{P}_A (a^\prime ) , 
\end{equation}
with 
\begin{equation}
\mu_X (x,x^\prime ) = \frac{1}{2} \left ( 1 + x x^\prime \frac{1}{\cos \theta} \right ) ,
\end{equation}
and
\begin{eqnarray}
\label{muZ}
& \mu_Z (1, 1 ) = \frac{\sin^2 \vartheta}{\sin \theta \sin ( 2 \vartheta - \theta) }, \quad 
\mu_Z (1, -1 ) = \frac{ - \cos^2 \vartheta}{\sin \theta \sin (2 \vartheta - \theta) }, & \nonumber \\
 & & \\
& \mu_Z (-1, 1 ) =  \frac{- \sin^2 ( \vartheta - \theta)}{\sin \theta \sin (2 \vartheta - \theta) }, \quad  
\mu_Z (-1, -1 ) =  \frac{\cos^2 (\vartheta - \theta)}{\sin \theta \sin (2 \vartheta - \theta) }, & \nonumber 
\end{eqnarray}

This inversion can be extended to the joint distribution with the aim of obtaining the impossible exact joint distribution for $X$ 
and $Z$ as
\begin{equation}
P (x,z) = \sum_{x^\prime, z^\prime} \mu_X (x, x^\prime )  \mu_Z (z, z^\prime ) \tilde{P} (x^\prime z^\prime) ,
\end{equation}
leading to 
\begin{equation}
P  (x,z)  = \frac{1}{4} \left [ 1+ x \delta (z) \langle X \rangle + z\langle Z \rangle \right ] ,
\end{equation}
where 
\begin{eqnarray}
\label{dz}
& \delta (1 ) =  \frac{\sin (2 \vartheta)}{\cos \theta \sin (2 \vartheta - \theta) } ,  & \nonumber \\
 & & \\
 &   \delta (-1 ) = \frac{\sin (2 \vartheta - 2 \theta)}{\cos \theta \sin (2 \vartheta - \theta ) }  .& \nonumber
 \end{eqnarray}
Since  $\delta(1)+\delta(-1) = 2$ we can appreciate that $P(x,z)$ provides the correct exact marginals (\ref{PZe}), (\ref{PXe})  for both 
observables. In particular extremely simple expressions are obtained in the case $\theta \rightarrow 0$ leading to 
\begin{equation}
\label{t0}
P (x,z) = \frac{1}{4} \left ( 1 + z \langle Z \rangle + x \langle X \rangle \right ) .
\end{equation}

The main conclusion is that for every state $| \psi \rangle$ we can suitable chose angles $\theta$ and $\vartheta$ so that $P(x,z)$ 
takes negative values. Focusing in the simplest case (\ref{t0}) the minimum value is 
\begin{equation}
P_\mathrm{min}  = \frac{1}{4} \left ( 1 - | \langle Z \rangle | - | \langle X \rangle | \right ) .
\end{equation}
We have that 
\begin{equation}
| \langle Z \rangle | + | \langle X \rangle |  \geq \langle Z \rangle^2 +  \langle X \rangle^2 = 1 ,
\end{equation}
where we have taken into account that we are working with pure states with $\langle Y \rangle =0$. Therefore  
$P_\mathrm{min}  < 0$ and $P(x,z)$ can no longer be probabilities in the standard sense. This does not mean 
that they are meaningless \cite{FE87}. For example they reveal that every state $| \psi \rangle$ is nonclassical. 

 \section{Joint  observation and data inversion: continuous phase}

An alternative and maybe more intuitive approach to interference or phase difference is provided by the phase states (\ref{ps}). With 
the help of them the joint distribution for the simultaneous observation of path and phase/interference is 

 \begin{widetext}
 \begin{equation}
 \tilde{P} (\phi,z)  =  \left | \langle \tilde{z} | \langle \phi |   \tilde{\psi} \rangle \right |^2,
 \end{equation}
\begin{equation}
 \tilde{P} (\phi,z)  = \frac{1}{2 \pi} \left [ \gamma_0 (z) + \gamma_X (z)  \cos \phi \langle X \rangle +   
 \gamma_X (z) \sin \phi \langle Y \rangle+ z \gamma_Z (z) \langle Z \rangle \right ] ,
 \end{equation}
\end{widetext}
with the same parameters $\gamma_A$ in Eq. (\ref{g}). In comparison with the discrete case the only essential difference is that  
$\phi$ assumes a continuous and $2 \pi$-periodic range of variation.

The marginal for $Z$ is the same in Eq. (\ref{PZa}), so the inversion is performed by the same matrix $\mu_Z$ in Eq. (\ref{muZ}). 
On the other hand, the marginal for the phase is 
 \begin{equation}
 \tilde{P}_\Phi (\phi) =  \frac{1}{2 \pi} \left [1 +  \cos \theta \left ( \cos \phi \langle X \rangle +  \sin \phi \langle Y \rangle \right )  \right ] ,
 \end{equation}
that can be inverted as 
 \begin{equation}
P_\Phi (\phi) = \int d \phi^\prime \mu_\Phi (\phi, \phi^\prime ) \tilde{P}_\Phi (\phi^\prime ) , 
\end{equation}
with 
\begin{equation}
\mu_\Phi (\phi, \phi^\prime ) =  \frac{1}{2 \pi} \left [ 1 + \frac{2}{\cos \theta} \cos (\phi - \phi^\prime  ) \right ] . 
\end{equation}

When applying the $Z$ and $\Phi$ inversions  simultaneously to the joint distribution we get 
\begin{equation}
P  (\phi ,z)  = \frac{1}{4 \pi } \left [ 1+ \delta (z) \left ( \cos \phi \langle X \rangle +  \sin \phi \langle Y \rangle \right ) + z\langle Z \rangle \right ] ,
\end{equation}
for the same $\delta (z)$ in Eq. (\ref{dz}). In the limit $\theta \rightarrow 0$ it becomes
\begin{equation}
\label{tt0}
P  (x,z)  = \frac{1}{4 \pi } \left [ 1+ \cos \phi \langle X \rangle + \sin \phi \langle Y \rangle + z\langle Z \rangle \right ] .
\end{equation}

The conclusions about the lack of positivity are the same as obtained above. The minimum value in Eq. (\ref{tt0})  is 
\begin{equation}
P_\mathrm{min}  = \frac{1}{4 \pi} \left ( 1 - | \langle Z \rangle | - \sqrt{ \langle X \rangle^2 + \langle Y \rangle^2 } \right ) .
\end{equation}
Since
\begin{equation}
| \langle Z \rangle | + \sqrt{ \langle X \rangle^2 + \langle Y \rangle^2 }  \geq \langle Z \rangle^2 + \langle X \rangle^2 +  \langle Y \rangle^2 = 1 ,
\end{equation}
where the last equality holds for pure states, we get again $P_\mathrm{min}  <0$ and the same conclusion as above.

\bigskip

\section{Conclusions}

Most practical and meaningful observations in quantum and classical physics are indirect in the sense that the desired information is 
retrieved after a suitable data analysis. This idea allows us to approach the statistics of conjugate observables by removing the instrumental 
effects of their imperfect simultaneous measurement. In classical physics this protocol always works providing {\it bona fide} joint probabilities. 

Here is where relies the most significant difference between classical and quantum physics. This is that in quantum mechanics 
this protocol fails, say it becomes a kind of {\it ghost protocol}, as a clear quantum signature. 
 
\section*{ACKNOWLEDGMENTS}

A. L. acknowledges financial support from Spanish Ministerio de Econom\'ia y Competitividad 
Projects No. FIS2016-75199-P, and from the Comunidad Aut\'onoma de Madrid research  consortium 
QUITEMAD+ Grant No. S2013/ICE-2801.

\end{document}